\documentstyle[prl,aps,multicol,epsf]{revtex}

\begin{document}

\draft

\title{Direct observation of the charging of a 2D electron gas through
an incompressible strip in the quantum Hall regime}

\title{Determination of the Longitudinal Resistance of
Incompressible Strips through Imaging of Charge Motion}

\author{P.I. Glicofridis, G. Finkelstein\footnote{Present
address: Department of Physics, Duke University, Durham, NC
27708}, R.C. Ashoori}

\address{Department of Physics and Center for Materials Science and
Engineering, Massachusetts Institute of Technology, Cambridge, MA
02139}

\author{M. Shayegan}

\address{Department of Electrical Engineering, Princeton University,
Princeton, NJ 08544}

\maketitle

\begin{abstract}

Using charge accumulation imaging, we measure the charge flow
across an incompressible strip and follow its evolution with
magnetic field. The strip runs parallel to the edge of a gate
deposited on the sample and forms at positions where an exact
number of integer Landau levels is filled. An RC model of
charging fits the data well and enables us to determine the
longitudinal resistance of the strip. Surprisingly, we find that
the strip becomes more resistive as its width decreases.

\end{abstract}


\begin{multicols}{2}

\narrowtext

The edge state picture of transport in a 2-dimensional electron
gas in a strong magnetic field appears to explain several key
properties of the 2DEG in the quantum Hall effect (QHE) regime.
The intersection between the constant Fermi energy and the
Landau-level energy near the edge of the 2DEG determines the
spatial location of the edge states. In a self-consistent
electrostatic picture, edge channels are separated by
incompressible strips of integer Landau level
filling\cite{halper,efros-etc,Chklovskii}. Resulting models
involving alternating compressible and incompressible strips have
received considerable theoretical and experimental attention
\cite{hwang,takaoka,zhiten}. In particular, groups have employed
various local techniques \cite{Westervelt,wei-set,McCormick,Yacoby} 
to resolve the structure of these strips
\cite{wei-set,McCormick,Yacoby}. Subsurface Charge Accumulation
(SCA) imaging enables measurements of both the disorder potential and
imaging of low compressibility strips in the integer
QHE\cite{contour-strip}.

In this work, we use SCA to image and characterize the charging
properties of the incompressible strips that form in the 2DEG in
the presence of a density gradient. The electron gas charges by
pumping charge to it across an incompressible strip. We show here
that a simple RC charging model fits the data well. Using this
model we determine the strip resistance $R_{strip}$ as a function
of magnetic field and find surprisingly, that the strips become
more resistive as their widths decrease.

Our samples are made from a standard GaAs/AlGaAs
heterostructure with a carrier concentration of $1.5 \times
10^{11}$ $cm^{-2}$ and transport mobility of $1.5 \times 10^6$
$cm^2/V$ sec. The 2DEG forms at the GaAs-AlGaAs interface 80 nm
beneath the surface. We deposited a Cr gate on top 
of the sample, lithographically patterned in the form of a 
fingered grating (see schematic in Fig.~1). An ac excitation of 
6 mV rms at a frequency of 100 kHz is applied to both the 
gate and the 2DEG. Due to its capacitance to the grounded 
body of the scanning microscope, the 2D electron layer charges and
discharges in
response to the applied signal. By measuring the charge induced
on a sharp metal tip scanned $\approx 10$ nm above the sample
surface at a temperature of 0.35 K, we obtain a charging map of
the 2DEG. Both the signal in-phase with and lagging $90^\circ$
from the ac excitation are recorded using a lock-in amplifier.

Figure 1a presents a $12\times12$ $\mu m$ SCA image at zero
magnetic field. These images result from scanning the tip over an
ungated portion of the sample situated between two gate fingers
that run parallel to each other. The scan window is positioned so
that only one finger appears on the left. In the absence of
magnetic field, the 2DEG maintains high conductivity and thus
charges uniformly in phase with the applied ac excitation. As a
consequence, the signal lagging the excitation (Y-phase) is zero
everywhere and the corresponding image (not shown) displays no
contrast. The picture changes dramatically in applied magnetic
field. At 6.4 Tesla (close to filling factor $\nu=1$), 
charge flowing from the
gates, now penetrates the 2DEG only partially (Fig.~1c). In
addition, the Y-phase signal in Fig.~1d peaks at the boundary
between charging and non charging regions.

In this work, we do not apply any external bias voltage on the gate.
However, the electronic density under the metal (chromium) gate
is reduced by about $15\% $ from its bulk value due to the
difference in chemical potential between the metal and GaAs. As a
result of this density profile, at high magnetic fields an 
incompressible strip, running parallel to each finger, forms 
in regions where electrons completely fill a Landau level. 
In these areas $\sigma_{xx}\rightarrow 0$, and the strip prevents 
charge from flowing freely into the interior of the sample, exactly as
depicted in Figure 1c. The regions of the 2DEG with very low
conductivity give rise to a phase lagging SCA signal (Fig. 1d).
Decreasing the strength of the applied magnetic field reduces the
degeneracy of the Landau levels. The strip, as well as the
boundary that separates charging from non charging regions, moves
closer to the fingers where the density is lower. By following
this evolution with field, we determine that the 2DEG electron 
density is nearly constant between the gate fingers but decreases 
closer to the gate. We find a density gradient of $\approx 1.2\times
10^{10} cm^{-2} \mu m^{-1}$ close to the gate.  By moving
our scan window laterally along the fingers of the gate, we
have verified that neighboring strips may close on each other at
certain locations, forming closed loops with the 2DEG inside
electrically isolated from the rest of the electron layer.

\begin{figure}
\epsfxsize=\linewidth
\epsfbox{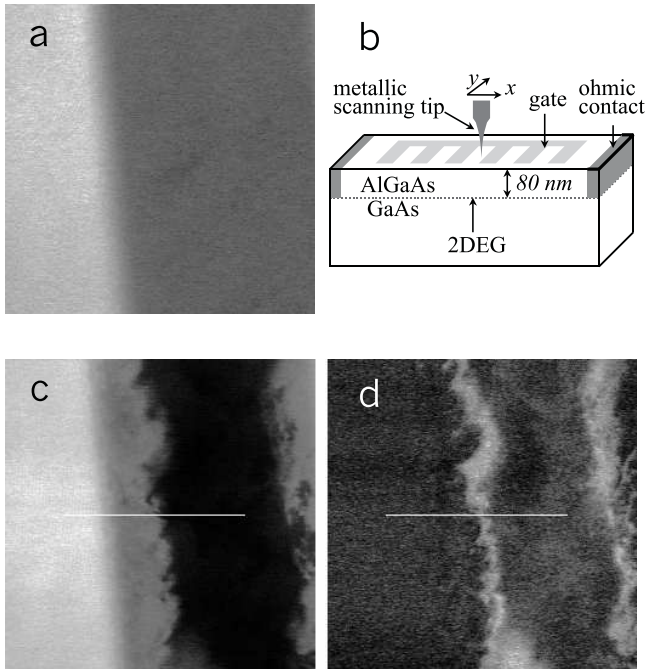}
\caption{\narrowtext
12x12 $\mu m$ SCA images: (a) B=0 T in-phase signal;
(c) 6.4 T in-phase and (d) 6.4 T out-of-phase. (b) sample schematic. 
The in-phase images (a and b) show the gate (bright region on the left 
side of the images), which is closer to the scanning probe than the 2DEG,
resulting in a larger capacitance signal. Incompressible strips at $\nu=1$
form close to the edge of the gate and separate regions of $\nu<1$
from those where $\nu>1$ (sample interior). Another gate sits at
$\approx 1\mu m$ off to the right of the images. White lines are
described in the captions for Fig. 3 and 4.}
\end{figure}

The work-function difference between the Pt-Ir scanning tip and
the GaAs surface results in an electric field that perturbs the
electronic density of the 2DEG in the vicinity of the tip. As
described elsewhere \cite{contour-strip}, we perform Kelvin probe
measurements and apply a bias voltage to the tip that nulls this
electric field. For values away from this nulling voltage, the
probe acts as an effective gate on the sample. Fig.~2 shows scans
of a 12x12 $\mu m$ region at $6.3$ Tesla for different tip
biases. With the local density under the tip reduced, the
condition for integer filling of the $\nu=1$ Landau level is
satisfied at a region with higher intrinsic electron density. As
a result, the location of the observed strip changes and shifts
towards the interior of the imaged region. The resulting density
suppression, though local, extends to $\approx 1 \mu m$ and
connects regions that would otherwise be separated by
incompressible fluid.

The shapes of areas with high out-of-phase signal in Fig.~2d
resemble sets of arcs. The centers of the arcs are located roughly
along the strip position measured at nulling voltage in Fig.~2a
suggesting the following non-local gating mechanism: bringing the
biased tip on-top of a high density region that does not charge
at nulling voltage may reduce the density and permit charging. In
this case, there must exist a path between the tip location and
the gate, where the filling factor is everywhere less than the
nearest integer. The density perturbation induced by the tip
decays sharply in the lateral direction. Crucial points along this
path are situated at the strip location as measured at nulling.
Indeed, in many cases the centers of the arcs such as seen in
Fig.~2d, point to the location of features in the unperturbed
images (see for example the feature in Fig.~2a marked by ``x").
We postulate that each arc indicates tip locations where the tip
influence on the ``weak link" in the strip is the same
\cite{Woodside}. On one side of the arc, the weak link is open
and the area under the tip charges fully. On the other side of
the arc, the weak link is closed and the area under the tip
becomes effectively disconnected from the gate (Fig.~2b).

We can counter changes in the position of the strip created with
the tip bias by properly tuning the magnetic field. At $\nu=1$,
the incompressible strips appear at the same location if we
change both magnetic field and tip voltage at a rate of dV/dB=0.5
V/T. At $\nu=2$ we find that dV/dB=1.0 V/T creates the same
effect. This demonstrates that density changes are compensated by
varying the number of electrons that each level can accommodate.
The voltage required for compensation should be proportional
to the filling factor, in agreement with the experiment. All
SCA images and measurements reported in this paper, other than 
those in Fig. 2 c and d, result from scans taken at nulled
tip-sample electric field.

We now describe the results of SCA measurements obtained by
scanning the tip along single lines shown schematically in Figure 1. 
The boundary between the gate finger and the bare
sample occurs at $x\approx 1.5 \mu m$ (distance of 1.5 $\mu m$
from the leftmost point of the line scan). The magnitude of the
resulting signal step reflects the difference in SCA signal
measured with the tip positioned above the metal gate ($0<x<1.5\mu
m$) and above a fully charging region of the 2DEG ($1.5 \mu m
<x<3.1 \mu m$). This observed step does not vary with magnetic
field (see Fig.~3), indicating no field dependence in the amount
of charge flowing in regions of the 2DEG adjacent to the gate.

\begin{figure}
\epsfxsize=\linewidth
\epsfbox{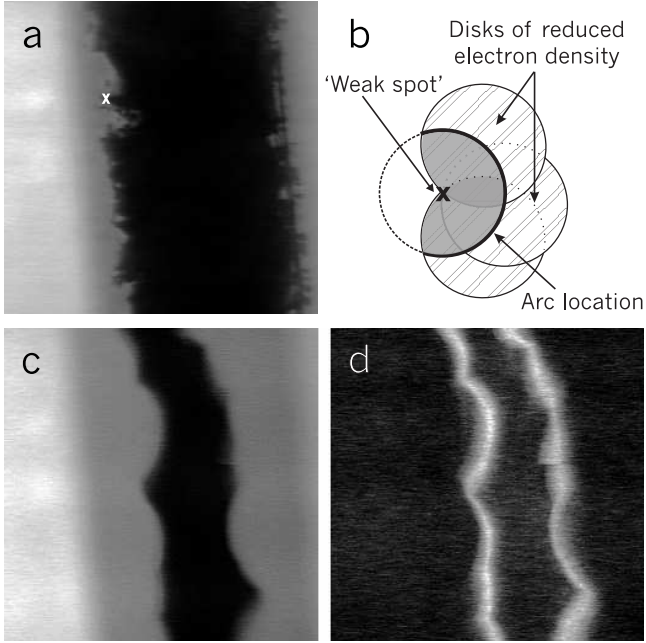}
\caption{\narrowtext
$12\times12$ $\mu m$ SCA images at B=6.3 T for
different biases applied on the tip with respect to the 2DEG. Tip
bias $V_{tip}$ is measured in Volts away from the nulling
voltage. (a) and (c) present in-phase signal at $V_{tip}=0$ 
(i.e. nulling) and $-1 V$, respectively. (d) is the out-of-phase 
signal corresponding to (c). The scan window is positioned so that 
two neighboring finger gates are imaged. The incompressible strips 
at $\nu=1$ are squeezed towards the two fingers as the density is 
raised with the tip. (b) shows schematically the origin of the arc 
shaped features observed in (c) and (d).}
\end{figure}

At zero magnetic field as well as at fields away from the quantum
Hall plateaus, the 2DEG charges nearly as a classical 2D metal.
This situation changes upon tuning the field close to the integer
Hall plateaus. The leftmost sequence (in-phase signal) of line
scans in Fig.~3 shows the appearance and evolution of another SCA
signal step at fields around 6.4~T. The decrease in signal in the
region between neighboring gate fingers results from the
formation of incompressible strips at $\nu=1$, running parallel
to the gate. We point out that this region in the interior,
bounded by incompressible strips on both sides, has $\nu>1$ and
high conductivity ($\sigma_{xx}$).

To understand the origin of the $90^{\circ}$ lagging SCA signal
(curves on the right of Fig.~3), we consider an RC charging model
for our 2DEG and assume that the interior of the sample charges
through a strip characterized by resistance R while the tip is
coupled to the 2DEG via a capacitance C (see inset in Fig.~4). As
discussed below, this tip-to-sample capacitance totally dominates
over the self-capacitance of the ``isolated" $\nu>1$ higher
conductivity region. The resulting in-phase signal measured at
the tip, is proportional to $1/(1+( \omega RC)^2)$ while the
out-of-phase evolves as ~$\omega RC/(1+(\omega RC)^2)$. The
X-signal drops steadily to zero as the resistance of the path 
leading to the region under the tip increases, while the Y-signal 
goes through a peak near the roll-off frequency of $1/(RC)$. 
The presence of the out-of-phase signal indicates that there is 
not enough time to fully charge the 2DEG during one complete 
cycle of the excitation.

\begin{figure}
\epsfxsize=\linewidth
\epsfbox{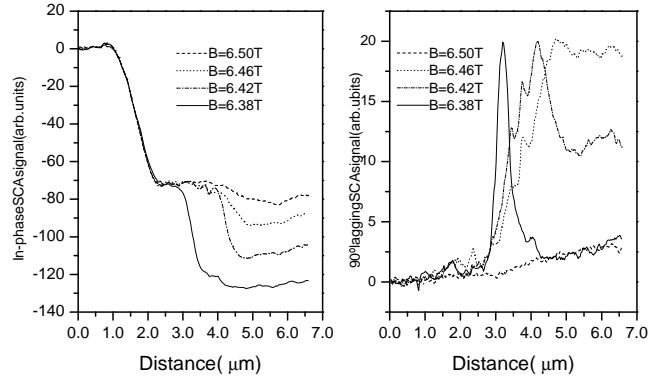}
\caption{\narrowtext
Frequency dependence of the measured capacitance
signal at 6.52 T, as extracted from the traces in the left inset.
The solid curve is an RC model fit to the data points. Left Inset:
In-phase SCA signal from line scans at 10, 30 and 100 kHz. Thick
bottom line is frequency independent and corresponds to the
sensitivity calibration trace measured at 6.44 T. Data are
measured at a different sample location from the one corresponding
to Fig.~3. Right Inset: Degree of charging of the 2DEG measured
for different tip-sample separations at 1, 10, 50, and 100 kHz.
Data points are derived from line scans at 6.40 Tesla at the same
location where measurements for Fig.~3 were taken (solid line in
Fig. 1 c and d). Reduction of $C_{g}$ due to the increase in $d$ 
leads to faster charging of the electron layer.}
\end{figure}

Figure 3 shows that the formation of the strip on the high field 
side of the Hall plateau (B=6.5 T) begins to impede charging of the 
interior $\nu>1$ region of the sample (Fig. 3, distance $ > 4 \mu$). 
At 6.46 and 6.42 T, we observe a somewhat higher strip resistance, 
but the interior can still be partially charged during the ac excitation
cycle, as evidenced by a nonzero phase lagging signal. With decreasing
field, the in-phase and phase lagging signals in the interior $\nu>1$
region further drop and eventually saturate beyond B=6.38 T. Further
reduction of the field simply pushes the strip closer to the edge of the
gate and leaves magnitude of the signals in the interior unchanged. 
We conclude that the resistance across the strip steadily grows as the
strip moves closer to the gate, and at B=6.38 T the strip becomes
practically impenetrable 
to charge flowing from the gate for excitation frequency of 100 kHz.
The density gradient is steeper closer to the gate and 
hence the incompressible strips formed there should be narrower
\cite{Chklovskii,contour-strip}. Therefore, the effective resistance 
for charging across the strip, counterintuitively, is  greater for a
narrower strip.

The crude RC model used so far dictates that the measured signal
also depends upon the excitation frequency. Lowering the
measurement frequency should enhance the charging of the 2DEG
through the strip provided that R and C remain constant. We show
data consistent with this expectation in the left inset of Fig.~4.
A line scan at 6.52T (in a different location from those in
Fig.~3), is measured at different frequencies. As expected, the
2DEG charges more fully at lower frequencies, as evidenced by the
increased in-phase signal in the $\nu>1$ region. The lower trace
in that inset corresponds to a magnetic field of 6.44 T where the
strip allows essentially no charge to cross it. In contrast to
line scans taken at slightly higher fields, this trace remains
unchanged even at the lowest experimental frequencies.

\begin{figure}
\epsfxsize=\linewidth
\epsfbox{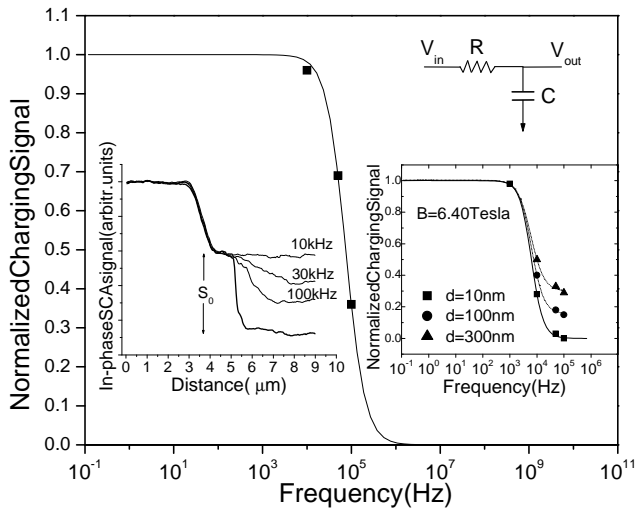}
\caption{\narrowtext
Evolution of scans taken along the solid white 
line of Figure 1 with magnetic field. Note the different vertical 
scales for the in-phase and out-of-phase curves. Charging of the 
sample interior becomes less efficient as the field is lowered. 
The strip is most resistive at 6.38 T where the Y-phase signal 
goes through a peak and returns again to almost zero level.}
\end{figure}

We use the corresponding step $S_{0}$ in the in-phase signal 
between the
fully charged and totally non-charged regions that the strip
separates (bold trace in the left inset of Fig.~4), to calibrate
the sensitivity of our measurement to full charging. We
can therefore determine precisely the degree of charging of the
2DEG in the region between the strips at different fields for a
range of frequencies and express it as a fraction $q$ of $S_{0}$.
Figure 4 plots $q$ against frequency using data from the curves
in the left inset. The smooth curve fits the data points using the RC
model discussed earlier and works remarkably well. We have
performed extensive measurements at several different locations
within the sample and find consistent agreement with the RC model.

We can now extract directly the resistance, R, of the strip by
assuming that the value for C is the known geometric capacitance
$C_{g}$ between the tip and the 2DEG. We justify this assumption
below in an examination of the effects of changing the tip-sample
separation. $C_{g}$ is determined by measuring the signal change
at zero magnetic field, as we move the tip vertically from
`infinity' to within 10 nm from the sample surface, yielding
$C_{g}\simeq$ 0.5 fF. In addition, since strips form closed loops
in our experiment, we are able to follow them to their full
extent and hence measure exactly their total length. For the fit
in Fig.~4 we obtain $RC=2\times10^{-6}$ s and deduce that at 6.52
T ($\nu=1$), $R_{strip}\approx 100$ M$\Omega$/$\mu m$. This
resistance increases by 1 order of magnitude to $\approx 1000$
M$\Omega$/$\mu m$ at B=6.48 T. We find that at a field of 6.44 T,
the strip resistivity rises beyond our measurement limit of $\sim
10000$ M$\Omega$/$\mu m$. The results reproduce consistently
at different locations throughout the sample.

We have also measured the strip resitances at $\nu=$ 2 and 4.
The strip resistivity is $\approx 10$ M$\Omega$/$\mu m$ at B=3.26 T 
and $\approx 400$ M$\Omega$/$\mu m$ at 3.24 T (corresponding to 
$\nu=2$). For B=1.60 T ($\nu=4$), we get $R_{strip}\sim 1$
~M$\Omega$/$\mu m$. This result follows naturally, since at $\nu=4$ 
the energy gap $\hbar\omega_{c}$ is smaller and the strips should 
be narrower \cite{Chklovskii}. Temperature dependent
measurements of the strip resistivity, could elucidate the
mechanism responsible for charge penetrating the incompressible
strip (i.e. electron tunneling or hopping). However, these
measurements prove difficult with our apparatus.

The effects of changing tip height are depicted in the right
inset of Fig.~4. The corresponding line scans are performed at
6.40 T at the same location as the data set in Fig.~3. The
capacitance data points are extracted at the same frequencies for
3 different tip-sample separations, $d=10 nm$, $100 nm$ and $300
nm$. Their strong dependence of charging rates on $d$ implies that
the geometric capacitance $C_{g}$ between the probe and the sample
dominates over the self-capacitance of the region of the 2DEG
located to the right of the strip. We obtained similar results
for lower field values up to the point where the strip becomes
totally impenetrable. Indeed, $C_{self}\sim \varepsilon_{0} a
\sim 1fF$, with $a$ being a characteristic length scale of the
2DEG region enclosed by the strip ($\sim 10 \mu m$), is about one
order of magnitude less than the value calculated for $C_{g}$ in
our experiment. Consequently, $C_{g}$ seems to be the relevant
parameter in our model and its known value may be used in
determining the strip resistance from the product $RC_{g}$.  From
Fig.~4 we also note that, as expected, the 2DEG charges faster as
$d$ increases and $C_{g}$ is reduced.

We would like to thank L.S. Levitov for helpful discussions. This
work is supported by the Office of Naval Research, the National
Science Foundation DMR and the Center for Materials Science and
Engineering at MIT.

\end{multicols}

\end{document}